\documentclass[aps,twocolumn,amssymb,showpacs]{revtex4}

\usepackage{epsfig}
\newcommand{\Def}{\newcommand}
\Def{\bq}{\begin{equation}}
\Def{\eq}{\end{equation}}
\Def{\bQ}{\begin{eqnarray}}
\Def{\eQ}{\end{eqnarray}}

\begin{document}
\title{Control of chaotic transport in Hamiltonian systems}
\author{Guido Ciraolo}
\affiliation{Facolt\`a di Ingegneria, Universit\`a di Firenze,
via S. Marta, I-50129 Firenze, Italy, and I.N.F.M. UdR Firenze}
\author{Cristel Chandre}
\affiliation{CPT-CNRS, Luminy Case 907, F-13288 Marseille Cedex 9,
France}
\author{Ricardo Lima}
\affiliation{CPT-CNRS, Luminy Case 907, F-13288 Marseille Cedex 9,
France}
\author{Marco Pettini}
\affiliation{Istituto Nazionale di Astrofisica, Osservatorio
Astrofisico di Arcetri, Largo Enrico Fermi 5, I-50125 Firenze, Italy,
and I.N.F.M. UdR Firenze}
\author{Michel Vittot}
\affiliation{CPT-CNRS, Luminy Case 907, F-13288 Marseille Cedex 9,
France}
\author{Charles Figarella}
\affiliation{Association Euratom-CEA, DRFC/DSM/CEA, CEA Cadarache,
F-13108 St. Paul-lez-Durance Cedex, France}
\author{Philippe Ghendrih}
\affiliation{Association Euratom-CEA, DRFC/DSM/CEA, CEA Cadarache,
F-13108 St. Paul-lez-Durance Cedex, France}
\date{\today}

\begin{abstract}
It is shown that a relevant control of Hamiltonian chaos is possible through
suitable small perturbations whose form can be explicitly computed. In
particular, it is possible to control (reduce) the
chaotic diffusion in the phase space of a Hamiltonian system with
$1.5$ degrees of freedom which models the diffusion of charged 
test particles in a
``turbulent'' electric field across the confining magnetic field in
controlled thermonuclear fusion devices. Though still far from
practical applications, this result suggests that some strategy to
control turbulent transport in magnetized plasmas, in particular
tokamaks, is conceivable.
\end{abstract}

\maketitle
Transport induced by chaotic motion is now a
standard framework to analyze the properties of numerous systems.
Since chaos  can be harmful in several contexts, during the last decade 
or so, much attention has been paid to the so-called topic of
{\it chaos control}.
Here the meaning of {\em control} is that one aims
at reducing or suppressing chaos inducing a relevant change
in the transport properties, by means of a small perturbation 
(either open-loop 
or closed-loop control of dissipative systems \cite{limapet,review}) 
so that the original structure of
the system under investigation is substantially kept unaltered.
Control of {\em chaotic transport} properties still remains an open issue
with considerable applications.\\ \indent In the case of dissipative systems, 
an efficient strategy of control works by stabilizing unstable 
periodic orbits, 
where the dynamics is eventually attracted.
Hamiltonian description of microscopic dynamics usually involves
a large number of particles. However, methods based on
targeting and locking to islands of regular motions in a 
``chaotic sea'' are of no practical use in control when
dealing simultaneously with a large number of unknown trajectories.
Therefore, the only hope
seems to look for {\it small perturbations}, if any,
making the system integrable or closer to integrable.
In what follows we show that it is actually possible
to control Hamiltonian chaos by preserving the Hamiltonian 
structure.\\ \indent Chaotic transport  of particles advected by a turbulent
electric field with a strong magnetic field
is associated with Hamiltonian dynamical systems under the 
${\bf E \times B}$ guiding center approximation~\cite{nota}.
Although it has been shown that the
$\bf E \times B$ drift motion of the so-called guiding center can
lead to a diffusive transport in a fairly good agreement with the
experimental counterparts \cite{marc88,marctur}, it is clear that such an
analysis is only a first step in the investigation and
understanding of turbulent plasma transport.
The control of
transport in magnetically confined plasmas is of major importance
in the long way to achieve 
controlled thermonuclear fusion. Two major mechanisms
have been proposed for such a turbulent transport, transport
governed by the fluctuations of the magnetic field and transport
governed by fluctuations of the electric field. There is presently
a large consensus to consider, at low plasma pressure, that the
latter mechanism  agrees with experimental
evidence \cite{Scott}. In the area of
transport of trace impurities, i.e. that are sufficiently
diluted so as not to modify the electric field pattern, 
the present model should be the
exact transport model. Even for this very restricted case, control
of chaotic transport would be very relevant for the
thermonuclear fusion program.
The possibility of reducing and even suppressing chaos
combined with the empirically found states of improved confinement in
tokamaks,
suggest to investigate the possibility to devise
a strategy of control of chaotic transport through
some smart perturbation acting at the microscopic level of charged
particle motions.\\ \indent First, we briefly
describe the Hamiltonian with $1.5$ degrees of freedom modeling 
the ${\bf E}\times{\bf B}$ motion of charged test particles in a ``spatially
turbulent'' electric field. Then we formulate the problem of control
and analytically derive the partial control term for a Hamiltonian describing 
the motion of these test particles.
Finally, we report the numerical evidence of the effectiveness 
of the method.
Let us begin by describing the model whose dynamics we want to
control. In the guiding center approximation, the equations of motion
of charged particles in the presence of a strong toroidal magnetic field
and of a nonstationary electric field are
\bq
{\dot{\bf x}}= \frac{d}{dt}{x \choose y}=\frac{c}{B^2}{\bf E}({\bf
x},t)\times {\bf B}= \frac{c}{B}{-\partial_y V (x,y,t)\choose
\partial_x V (x,y,t)} , \label{guidcent}
\eq
where $V$ is the electric potential, ${\bf E}=-{\bf \nabla} V$,
and ${\bf B}=B {\bf e_z}$. To define a model we choose
\bq
V ({\bf x},t)=\sum_{\bf k} {V_{\bf k}}\sin [{\bf k}\cdot {\bf x}
+\varphi_{\bf k}-\omega ({\bf k})t],
\eq
where $\varphi_{\bf k}$ are random phases and the set of 
$V_{\bf k}$'s decreases as
a given function of $|{\bf k}|$, in agreement with experimental data
\cite{anormal_exp}.
In principle, one should use for $\omega ({\bf k})$ the dispersion
relation for electric drift waves (which are thought to be
responsible for the observed turbulence) with a frequency broadening
for each ${\bf k}$ in order to model the experimentally observed
spectrum $S({\bf k},\omega)$. Unfortunately this would be prohibitive
from a computational point of view, therefore one is led to simplify
the model drastically by choosing $\omega({\bf k})=\omega_0$ constant
and the phases $\varphi _{\bf k}$ at
random to reproduce a turbulent field
(with the reasonable hope that the properties of the
realization thus obtained are not significantly different from their
average). In addition we take for $|{V_{\bf k}}|$ a power law in
$|{\bf k}|$ to reproduce the spatial spectral characteristics of the
experimental $S({\bf k})$, see Ref.~\cite{anormal_exp}. Thus
we consider the following explicit form for the electric potential
\bq
V (x,y,t) =\frac{a}{2\pi}\sum_{m,n=1\atop{n^2+m^2\le N^2}}^N
\frac{\sin \left[\frac{2\pi}{L}(nx + my) +
\varphi_{nm} -\omega_0 t \right]} {(n^2+m^2)^{3/2}}~.
\label{potential}
\eq
By rescaling space and time, we can always assume that $L=1$ and $\omega_0=1$.
In what follows, we choose $N=25$. 
The spatial coordinates $x$ and $y$ play the role of the canonically
conjugate variables. We extend the phase space $(x,y)$ into
$(x,y,E,\tau)$ where the new dynamical variable $\tau$ evolves as 
$\tau(t)=t+\tau(0)$ and $E$ is its canonical conjugate. We absorb
the constant $c/B$ of Eq.~(\ref{guidcent}) in the amplitude $a$, so that
we can assume that $a$ is small  when $B$ is large. The 
autonomous Hamiltonian of the model is
\bq
H(x,y,E,\tau) = E + V(x,y,\tau)\label{hamilton}~.
\eq
The equations of motion are
\bq
{\dot x}=\frac{\partial H}{\partial y} =
\frac{\partial V}{\partial y}~,\ \ \ \ \ \ \ {\dot y}=-\frac{\partial
H}{\partial x} = -\frac{\partial V}{\partial x}~,\ \ \ \ \ \ \ {\dot \tau}= 1~,
\label{Hequations}
\eq
and $E$ is given by taking $H$ constant along the trajectories.
Thus, for small values of $a$, Hamiltonian 
(\ref{hamilton}) is in the form $H=H_0+\epsilon V$, 
that is an integrable 
Hamiltonian $H_0$ (with action-angle variables)
plus a small perturbation $\epsilon V$. For simplicity we 
assume that the average
of $V$ over the angles is zero. Otherwise similar calculations
following these lines can be done.
The problem of control in Hamiltonian systems is to find a small 
perturbation term $f$ such that $H+f$ is integrable. In this article, 
we are interested in finding a {\em partial} control term
$f_2$ of order $\epsilon^2$
such that the Hamiltonian given by $H_c=H_0+\epsilon V+\epsilon^2 f_2$ 
is closer to integrability, i.e. such that $H_c$ is canonically 
conjugate to $H_0+O(\epsilon^3)$. We perform a Lie transform
on $H_c$, generated by a function $S$:
\bq
H^{\prime}_c=e^{\epsilon \hat S} H_c\equiv H_c+\epsilon\{S,H_c\}
+\frac{\epsilon^2}{2}\{S,\{S,H_c\}\}+....,
\eq
where $\{\cdot~,\cdot\}$ is the Poisson bracket and the operator
$\hat S$ is acting on $H$ as $\hat S H = \{S,H\}$.
An expansion in power series in $\epsilon$ of $H^{\prime}_c$ gives
\bQ
H^{\prime}_c&=&H_0+\epsilon\left[\{S,H_0\}+V\right]\nonumber\\
&&+\epsilon^2\left[f_2+\{S,V\}+
\frac{1}{2}\{S,\{S,H_0\}\}\right]+O(\epsilon^3)~.\nonumber\\
\eQ
The generating function $S$ is chosen such that
\bq
\{S,H_0\}+V=0~,
\label{gen_fun}
\eq
provided that this equation has a solution.
The control term $f_2$ given by the cancellation of order $\epsilon^2$ terms
$[$and using Eq.~(\ref{gen_fun})$]$,
\bq
f_2=-\frac{1}{2}\{S,V\},
\eq
satisfies the required condition that $H_c$ is canonically conjugate
to $H_0$ up to order $\epsilon^3$ terms. We notice that by adding 
higher order terms in $\epsilon$ in the 
control term, one can build $f$ such that 
$H_c=H+\epsilon^2 f$ is integrable
for sufficiently
small $\epsilon$ (see Ref.~\cite{michel} for more details and for 
a general formulation and theorem).\\ In the 
case we consider, $H_0=E$
and Eq.~(\ref{gen_fun}) becomes
\bq
-\frac{\partial S}{\partial t}+V=0~,
\eq
and so $S$ is one primitive in time of $V$.
We choose the one with zero time average.
For the model (\ref{potential}), the generating function $S$ is
\bq
S (x,y,\tau) =\frac{a}{2\pi}\sum_{m,n=1\atop n^2+m^2\le N^2}^N \frac{\cos \left[2\pi(nx + my) +
\varphi_{nm} -\tau \right]} {(n^2+m^2)^{3/2}}~,
\label{S}
\eq
and the computation of $f_2$ gives
\bQ
&&f_{2}(x,y,\tau)=
\frac{a^2}{8\pi}\sum_{n_1,m_1, n_2,m_2} \frac{n_2 m_1 - n_1 m_2}{ 
 (n_1^2+m_1^2)^{3/2} (n_2^2+m_2^2)^{3/2}}  \nonumber\\
&&\times\sin \bigl[ 2\pi \bigl[ (n_1-n_2) x + (m_1-m_2) y\bigr] +
\varphi_{n_1 m_1} - \varphi_{n_2 m_2} \bigr]~.  \nonumber
\label{approx-control}\\
\eQ
We note that for the particular model
(\ref{potential}), the partial control term
$f_2$ is independent of time.\\ \indent With the aid of numerical 
simulations (see Ref.~\cite{marc88}
for more details on the numerics), we check the effectiveness of
the above control by comparing the diffusion properties of 
particle trajectories obtained from Hamiltonian (\ref{potential}) and
from the same Hamiltonian with the control term (\ref{approx-control}).
Figures 1 and 2 show the Poincar\'e surfaces of section of two
trajectories issued from the same initial conditions computed without
and with the control term respectively. Similar pictures are obtained 
for many other randomly chosen initial conditions.
 A clear evidence is found for a
relevant reduction of the diffusion in presence of the control term
(\ref{approx-control}).
\begin{figure}
\epsfig{file=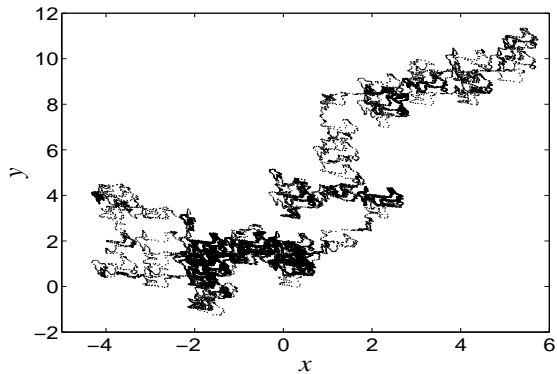,width=7.5cm,height=5cm}
\caption{Poincar\'e surface of section of a trajectory obtained for
Hamiltonian (\ref{potential}) assuming $a=0.8$.}
\label{figure1}
\end{figure}
\begin{figure}
\epsfig{file=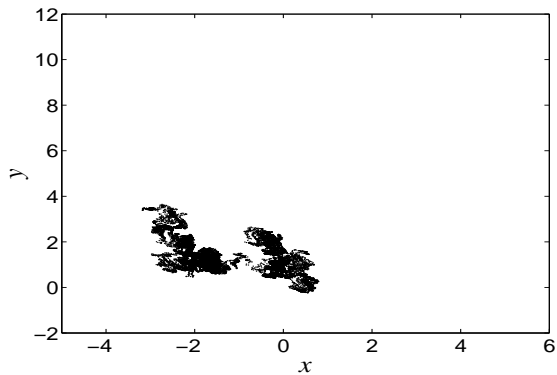,width=7.5cm,height=5.0cm}
\caption{Poincar\'e surface of section of a trajectory obtained for
the same initial condition as in Fig.1 and adding the control term 
(\ref{approx-control}) to Hamiltonian (\ref{potential}) .}
\label{figure2}
\end{figure}
\\ \indent In order to study the diffusion properties of the system, we have
considered a set of $\mathcal M$ particles (of order $100$) 
uniformly distributed at
random in the domain $0\leq x,y\leq 1$ at $t=0$. We have computed the
mean square displacement $\langle r^2 (t) \rangle$ as a function of
time
\bq
\langle r^2 (t) \rangle = \frac{1}{\mathcal M} \sum_{i=1}^{\mathcal M}
{|{\bf x}_i(t) - {\bf x}_i(0)|}^2
\eq
where ${\bf x}_i(t),~i=1,\dots,\mathcal M$ is the position of the
$i$-th particle at time $t$ obtained by integrating Eq.~(\ref{Hequations}) 
with initial condition ${\bf x}_i(0)$.
Figure 3 shows $\langle r^2 (t) \rangle$ for three different values of
$a$.
\begin{figure}
\epsfig{file=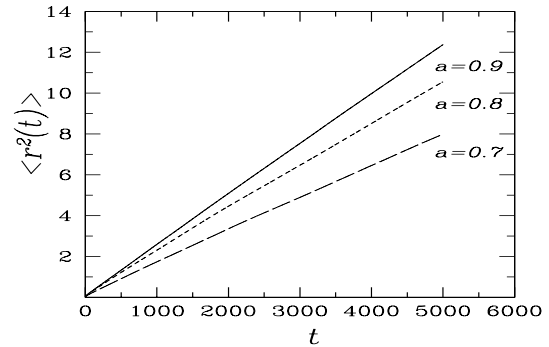,width=7.5cm,height=5.0cm}
\caption{Mean square displacement $\langle r^2 (t) \rangle$ versus
time $t$ obtained for  Hamiltonian
(\ref{potential}) with three different values of $a=0.7,0.8,0.9$.}
\label{figure3}
\end{figure}
For the range of parameters we consider, the behavior 
of $\langle r^2 (t)\rangle$ is always found to be 
linear in time for $t$ large enough. The
corresponding diffusion coefficient is defined as
\[
D= \lim_{t \rightarrow\infty}{{\langle r^2 (t) \rangle} \over t}~.
\]
Figure $4$ shows the values of $D$ as a function of $a$ with and without 
control term. It clearly shows a significant decrease of the diffusion 
coefficient when the control term is added.
As expected, the action of the control term gets weaker as $a$ is
increased towards the strongly chaotic phase.
\begin{figure}
\epsfig{file=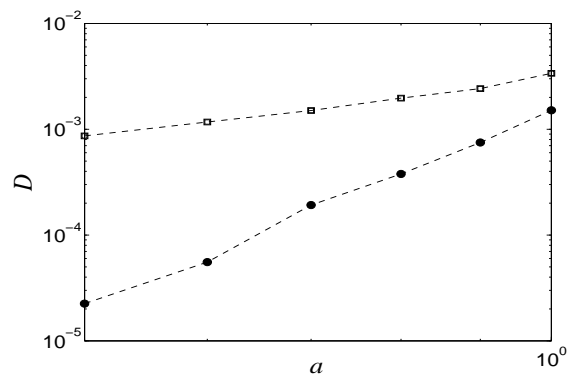,width=7.5cm,height=5.0cm}
\caption{Diffusion coefficient $D$ versus $a$ in log-log scales
obtained for Hamiltonian (\ref{potential}) (open squares) and Hamiltonian
(\ref{potential}) plus control term (\ref{approx-control}) (full
circles).}
\label{figure4}
\end{figure}
\\ \indent We check the robustness of the control by replacing $f_2$
by $\delta\cdot f_2$ and varying the parameter $\delta$ away from its
reference value $\delta =1$. Figure 5 shows that increasing
or decreasing $\delta$ from $\delta=1$ result
in a loss of efficiency of the control.
The fact that a larger
perturbation term ($\delta > 1$) does not work better than the one with
$\delta =1$  means that the perturbation is ``smart'' 
and that it is not a ``brute force'' effect.
\begin{figure}
\epsfig{file=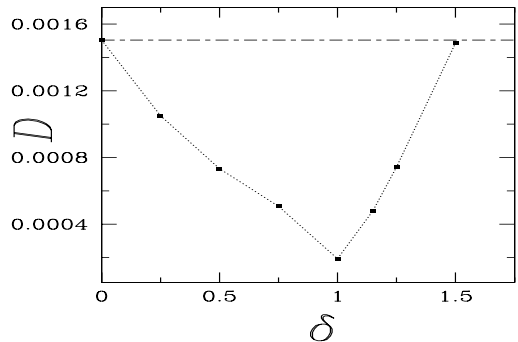,width=7.5cm,height=5.0cm}
\caption{Diffusion coefficient $D$ versus $\delta$ magnitude of the control term
(\ref{approx-control}) for a fixed value of $a = 0.7$}
\label{figure5}
\end{figure}

Let us define the {\em horizontal step size} (resp.~vertical step size) 
as the distance covered by the test particle between two successive sign
reversals of the horizontal (resp.~vertical) component of the drift
velocity. The effect of the control is analyzed in terms
of the Probability Distribution Function (PDF) of step sizes.
Following test particle trajectories for a large number of initial
conditions, with and without control, leads to the PDFs plotted in
Fig.~$6$. A marked reduction of the PDF is observed at large step
sizes with control relatively to the not controlled case.
Conversely, an increase is found for the smaller step sizes. 
\begin{figure}
\epsfig{file=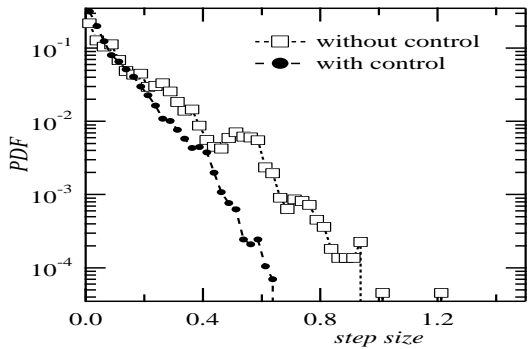,width=7.5cm,height=5.0cm}
\caption{PDF of the magnitude of the horizontal step size with
and without the control term.}
\label{figure6}
\end{figure}
The
control quenches the large steps, typically larger 
than $0.5$.\\ \indent In order
to measure the relative magnitude between Hamiltonian
(\ref{potential}) and $f_2$, we have numerically computed their mean
squared values:
\bq
\sqrt{\frac{\langle f_{2}^{2}\rangle} 
 {\langle V^2\rangle}} \approx 0.13 a~~,
\eq
which means that the control term can be considered a small
perturbative term (when $a<1$).

So to conclude this work, we have provided an effective new strategy to
control the chaotic diffusion in Hamiltonian dynamics using  small
perturbations. Since the formula
of the control term is explicit, we are able to compare the dynamics
without and with control in a simplified model, describing anomalous
electric transport in magnetized plasmas. Even though we use a
rather simplified model to describe chaotic transport of charged
particles in fusion plasmas, our result makes conceivable that to
apply some smart perturbation can lead to a relevant reduction of
the turbulent losses of energy and particles in tokamaks.


\begin{thebibliography}{99}
\bibitem{nota}Interestlingly enough, the dynamics of test particles
streaming along a field line in a fusion device are also
governed by Hamiltonian dynamics, the Hamiltonian being either 
the poloidal or toroidal magnetic flux. The principle of control of
stochastic transport would apply equally well to magnetic turbulence.
\end{thebibliography}
\end{document}